\documentclass[usenatbib]{mn2e}
\usepackage{graphicx, amssymb, aas_macros, natbib}
\usepackage{epsfig} % , deluxetable}
\usepackage{mathrsfs,amssymb}
\usepackage{graphicx,latexsym}
\newcommand{\Msun}{\mbox{M$_\odot$}}
\newcommand{\msun}{\mbox{M$_\odot$}}
\newcommand{\mvir}{\mbox{M$_{\rm vir}$}}

\newcommand{\Mstar}{\mbox{M$_\star$}}

\begin{document}

\title[Local Group constraints on high-redshift stellar mass functions]{Push it to the limit: Local Group constraints on high-redshift stellar mass functions for $M_{\star} \ge 10^5\,M_{\odot}$}

\author[Graus et al.]{Andrew S. Graus$^1$\thanks{$\!\!$e-mail: agraus@uci.edu}, 
James S. Bullock$^1$, Michael Boylan-Kolchin$^2$, Daniel R. Weisz$^3$$^4$ \\
$\!\!^1$Center for Cosmology, Department of Physics and Astronomy, 
  4129 Reines Hall, University of California Irvine, CA 92697, USA \\
$\!\!^2$ Department of Astronomy, The University of Texas at Austin, 2515 Speedway, Stop C1400, Austin, TX 78712, USA\\
$\!\!^3$Astronomy Department, Box 351580, University of Washington, Seattle, WA, USA\\
$\!\!^4$Hubble Fellow\\}
\maketitle

%==============================================================================
\begin{abstract}
We constrain the evolution of the galaxy stellar mass function from 2 $< z <$ 5 for galaxies with stellar masses as low as $10^{5}$ $\Msun$ by combining star formation histories of Milky Way satellite galaxies derived from deep Hubble Space Telescope observations with merger trees from the {\tt ELVIS} suite of N-body simulations. This approach extends our understanding more than two orders of magnitude lower in stellar mass than is currently possible by direct imaging.  We find the faint end slopes of the mass functions to be $\alpha$ = $−1.42^{+0.07}_{-0.05}$ at $z=2$ and $\alpha$ = $−1.57^{+0.06}_{-0.06}$ at $z=5$, and show the slope only weakly evolves from $z=5$ to $z=0$. Our findings are in stark contrast to a number of direct detection studies that suggest slopes as steep as $\alpha$ = -1.9 at these epochs. Such a steep slope would result in an order of magnitude too many luminous Milky Way satellites in a mass regime that is observationally complete ($M_{\star} > 2\times10^5 \Msun$ at $z=0$). The most recent studies from ZFOURGE and CANDELS also suggest flatter faint end slopes that are consistent with our results, but with a lower degree of precision.  This work illustrates the strong connections between low and high-$z$ observations when viewed through the lens of $\Lambda$CDM numerical simulations.
\end{abstract}

\nocite{Song15}

%\dsp
%==============================================================================
\section{Introduction} \label{s:intro}

The evolution of the galaxy stellar mass function, from high-redshift to low, provides an important benchmark for self-consistent galaxy formation models set within the Lambda Cold Dark Matter  ($\Lambda$CDM) framework \citep[][]{White78,Somerville14,Vogelsberger14,Schaye15}. 
In particular, the low-mass slope of the stellar mass function (hereafter, $\alpha_{\rm faint}$) is a crucial gauge of how and when feedback becomes important in suppressing galaxy formation in small dark matter halos.  Indeed, many of the best models today predict that the galaxy stellar mass function should be much steeper at high redshift than low redshift, more faithfully tracing the dark matter halo mass function slope itself \citep[e.g.,][]{Genel14,Furlong15}, while others produce a flatter slope at high redshifts via strong feedback \citep[][]{Lu14}.  As we detail below, a survey of the observational literature over the last several years finds little agreement on whether the stellar mass function was much steeper at $z = 2-5$ than it is at $z=0$.   In what follows, we use local observations to inform this question, relying on the large, uniform sample of star formation histories (SFHs) that are now available for local dwarf satellite galaxies \citep{Weisz14} together with the {\tt ELVIS} N-body simulation suite of the Local Group \citep[][]{SGK14} to connect local galaxy properties to the global galaxy population over cosmic time.

Significant effort has been dedicated to characterizing the galactic stellar mass function (GSMF) and its evolution.  A good $z \sim 0$ reference point comes from \citet{Baldry12}, who have used data from the GAMA survey to characterize the GSMF down to $M_{\star} = 10^8 \msun$ and find $\alpha_{\rm faint} \simeq -1.47$. This result has been echoed by other recent work such as \citet{Moustakas13} whose result agrees quite well with the \citet{Baldry12} stellar mass function, although they do not quote a value for $\alpha_{\rm faint}$.

Characterizing the mass function at higher redshift is also a priority. However, the results have been divergent, despite a significant
effort.    For example,  \citet{PG08} and \citet{Marchesini09} both reported fairly flat slopes with $\alpha_{\rm faint} \simeq -1$ to $-1.2$ out beyond $z = 2$, while \citet{Gonzalez11}  found steeper low-mass slopes $\alpha_{\rm faint} \simeq  -1.4$ to $-1.6$ from redshift $z = 4-6$, consistent with little evolution from the $z \sim 0$ value.  Still a different conclusion was presented by \citet{Santini12}, who reported significant evolution, with a slope similar to the low-redshift value $\alpha_{\rm faint} \simeq -1.45$  at $z = 1$ but steepening to $\alpha_{\rm faint} \simeq -1.85$ at $z=3$. 

Recent work has explored the stellar mass function to lower stellar masses, but there is still no consensus on the value of $\alpha_{\rm faint}$ and its evolution.   \citet[][]{Ilbert13} report best-fit slopes $\alpha_{\rm faint} \simeq -1.45$ to $-1.6$ out to $z \simeq 1.5$ using UltraVISTA  and \citet{Tomczak14} use ZFOURGE/CANDELS and find similar best-fit values of $\alpha_{\rm faint} \simeq -1.5$ to $-1.6$ out to $z \simeq 3$ with no obvious evolution.  In contrast to this apparent lack of evolution, \citet{Duncan14} have used CANDELS and {\textit{Spitzer}} IRAC data in the GOODS-S field to quantify the GSMF and find quite steep slopes ($\alpha_{\rm faint} \simeq -1.75$ to $-1.9$) at $z = 4-5$.  \citet{Grazian15} have done similar work in the CANDELS/UDS, GOODS-South, and HUDF fields and prefer slightly shallower values ($\alpha_{\rm faint} \simeq -1.63$) over the same redshift range.  Most recently, \citet{Song15} have combined the CANDELS and HUDF fields with the deepest IRAC data to date to characterize the GSMF to an unprecedented depth of $M_\star \simeq 10^7$ $\Msun$ and report an even flatter slope at $z=4$, $\alpha_{\rm faint} = -1.53$, but see continued steepening to $\alpha_{\rm faint} = -2.45$ at $z = 8$.     

The idea that the stellar mass function might be getting steeper at early times is qualitatively consistent with the prevailing notion that cosmic reionization relies fundamentally on ionizing photons produced by a large number of faint galaxies \citep{alvarez2012, kuhlen2012a, Schultz2014, robertson2015, bouwens2015}.  Of course, a high comoving abundance of faint galaxies at early times requires that small dark matter halos are efficient at producing stars at this time \citep[e.g.,][]{Madau14}, an idea that is qualitatively at odds with what must happen in the $z = 0$ universe.   Specifically, if $\Lambda$CDM is correct, then the Milky Way should be surrounded by thousands of small dark matter subhalos ($\mvir \sim 10^{9} \msun$), far more than the number observed as dwarf galaxies \citep{Moore99,Klypin99}.  Solving the �missing satellites problem� requires the suppression of galaxy formation in small
halos today.  
%smallest halos, a regime where SNe feedback \citep{Dekel86,White91,Somerville99},  inefficient H$_2$ formation \citep{Robertson08,Kuhlen12,Jaacks13}, and UV- related feedback \citep{Efstathiou92,Bullock00,Okamoto08,Sawala15} are all expected to suppress dwarf galaxy formation.  
When (and if) small dark matter halos transition from being the sites of efficient star formation to less efficient star formation is a question that informs our ideas
about the origin of strong feedback in small halos \citep{Jaacks13,Wise14}.  Unfortunately, the luminosities of interest are unobservably faint ($M_{\rm UV} \sim -14$ at $z > 6$) with current telescopes, and likely will remain unobservable even in the era of $JWST$. 
 
An alternative way to investigate this question observationally is to use local observations as a time machine \citep{Weisz14b,MBK14,MBK15}, a possibility that is enabled by SFHs derived by deep color-magnitude diagrams from $HST$ \citep{Weisz14}.    These precise SFHs can be used to calculate UV luminosities for each galaxy at specific redshifts \citep{Weisz14b}, and demonstrate that the progenitors of many Milky Way's satellites (e.g., Fornax, Sculptor, Draco) had UV luminosities during the epoch of reionization that coincide well with the galaxies that seem to be required to maintain reionization at $z \sim 8$ \citep[][B15 hereafer]{MBK15}.  Moreover, by enumerating these galaxies today, and tracking back their expected abundances using N-body simulations,  \citet[][B14 hereafter]{MBK14} showed that the $z=8$ UV luminosity function (LF) must break to a shallower slope than the observed $\alpha \approx -2$ for galaxies fainter than $M_{\rm UV} \sim -14$.  The constraint was derived from the seemingly unavoidable prediction that there are hundreds of $z=0$ surviving descendants of reionization-era atomic cooling halos in the Local Group.  

This paper adopts a similar approach as in B14, but now applied to constrain stellar mass functions at
intermediate redshfits $z = 2-5$.    The upper limit of this redshift range is set by the age for reliable SFHs that can be achieved by isochrone fitting, which allows one to measure stellar mass older than $\sim 12.5$ Gyr (corresponding to $z \sim 5$) but with no finer precision \citep[e.g.,][]{Weisz14}.  The lower limit ($z=2$) is informed by our theoretical methodology.  Specifically, our method relies on studying the high-redshift progenitors of surviving subhalos in the simulations.  We associate surviving subhalos with dwarf satellite galaxies and treat their progenitors as typical field galaxies at earlier times.  In order to do this, we need to ensure that these systems were not satellites at the redshift of concern.  Most subhalos of Milky Way size hosts in our simulations are accreted after $z=2$ (see Section 2.1 below).  After this time, at redshifts $z \lesssim 2$, the SFHs of satellite galaxies are likely affected by environmental quenching \citep{Fillingham15,Wetzel15a,Rocha12}, and therefore are poor proxies for typical galaxies in the universe prior to this epoch.  By restricting our analysis to $z \geq 2$ we avoid this clear bias.

The paper is organized as follows, Section 2 gives a summary of the simulations and observational data used.  Section 3 presents in detail how we used the data to look for the effects of galaxy suppression at high redshift and presents the main results of this paper.  Section 4 discusses the implications of this work, and provides predictions for high redshift stellar mass functions. The conclusions are presented in section 5, along with possible caveats and alternative explanations. The cosmology of the {\tt ELVIS} suite and thus adopted for the analysis within this paper is WMAP-7 (\citealt{Larson11}: $\sigma_{8}$=0.801, $\Omega_{m}$ = 0.266, $\Omega_{\Lambda}$ = 0.734, $n_{s}$ = 0.963, and $\it{h}$ = 0.7)

%==============================================================================
\section{Data} \label{s:data}

\subsection{Simulation Data}

We make use of the {\tt ELVIS} suite \citep{SGK14} of N-body simulations, which is composed of 12 pairs of Milky Way-Andromeda analogs along with 24 mass-matched isolated systems for comparison. We use the 10 best-matched Milky Way-Andromeda analogs in this work (see Garrison-Kimmel et al. 2014 for details).  The simulations are complete to peak halo masses of $M_{\rm halo} = 6 \times 10^{7}$ $M_{\odot}$,  well below the mass range that can host known (classical) satellite galaxies of the Milky Way \citep[e.g.][]{MBK12}, which are the points of comparison in this paper.  Crucial to our analysis is the ability to connect bound subhalos at $z=0$ to their progenitors at higher redshift.  We do this using the merger trees provided in the public {\tt ELVIS} data base\footnote{http://localgroup.ps.uci.edu/elvis/}, which were constructed using the Rockstar \citep{Beh13} halo finder.

For our high redshift analysis we use only subhalos that survive as bound substructures at $z=0$ within 300 kpc of either host. We make no assumption as to which of the pairs on the {\tt ELVIS} catalogs would be the Milky Way or Andromeda.  This assumption is fair because the mass estimates of the Milky Way and Andromeda are the same within errors (\citealt{VDM12}; \citealt{MBK13}). 

One of our main assumptions in this work is that it is fair to treat the main progenitors of subhalos as typical galaxies prior to the time they first became subhalos. We know that classical dwarf satellite galaxies are more likely to be quenched than their counterparts in the field \citep{Mateo98} and we therefore want to avoid this bias in our analysis.~\footnote{Ultrafaint galaxies may likely be quenched, even in the field, owing to reionization \citep{Bullock00,Ricotti05}.}

Using the {\tt ELVIS} simulations, \citet{Wetzel15} found that $\sim 84 \%$ of subhalos in the relevant mass range were accreted after $z=2$ and $\sim 98 \%$ were accreted after $z=3$.  Our own independent analysis yields similar results \citep[see also][]{Fillingham15}.  This fact motivates our use of $z=2$ as the lower bound on the redshift range.  We note that this choice is also consistent with direct estimates for the accretion times of the classical Milky Way satellites from \citet{Rocha12},  who find that only one of the Milky Way satellites in our sample (Ursa Minor) is consistent with having been accreted before $z=2$, with an infall time between 8 and 11 Gyr ago.

\subsection{Observational Data}

\subsubsection{Milky Way Satellite Star Formation Histories}

The star SFHs of Milky Way satellites are taken from the data sets of \cite[][W14]{Weisz14} and \citet[][W13]{Weisz13}, which rely color magnitude diagrams from archival {\textit{HST}}/WFPC2 observations.  We refer the reader to \cite{Weisz14} and \citet{Dolphin02} for an in-depth discussion of the techniques used.   In what follows we assume that the SFHs are applicable to the entire galaxy, even though in many cases the CMDs were
derived from fields that do not cover the entire galaxy.  Instead, we use the normalized SFHs from W14 and W13 to the $z=0$ stellar mass of each Milky Way satellite using the values given in \cite{McConnachie12}, which were calculated based on integrated light, and assuming a mass to light ratio of one. 

One of our major concerns is that our comparison set be complete.   For this reason, we restrict ourselves to the 12 dwarfs currently within 300 kpc of the Milky Way and with $M_{\star}^ {\it{z}={\rm0}} > 2\times10^5 \, \Msun$:  Carina, Canes Venatici I, Draco, Fornax, Leo I, Leo II, LMC, Sagittarius, Sculptor, Sextans, SMC, and Ursa Minor. The SFHs in the W13/W14 data set include all of these bright Milky Way satellites except Sextans. In our fiducial analysis we include Sextans ($M_\star^{\it{z}={\rm0}} = {\rm 4.4 \times 10^5}$ $\Msun$) by assuming its normalized SFH is the same as that of Carina ($M_\star^{\it{z}={\rm0}} = {\rm 3.8 \times 10^5}$ $\Msun$).  We find that alternative choices for Sextans do not strongly alter our results.

%\citet{Wetzel15}, also investigated the assumption that few of our satellites fell into the Milky Way prior to z = 2.  Within the ELVIS %simulation the authors find that no present day satellites of the Milky Way or Andromeda were satellites already at reionization. This fraction %rises with redshift however, 95 \% of satellites are still outside of $R_{vir}$ at z = 3 and less than 68 \% at z = 2. Therefore, the further %beyond z = 2 we look, the more we corrected for satellite accretion onto the Milky Way. In order to use this data set to constrain the properties %of the high redshift universe we rely on two methods.

\begin{figure*}
\centering
\includegraphics[width=7.5in, height=7.5in, trim = .8in 0 0 0 0]{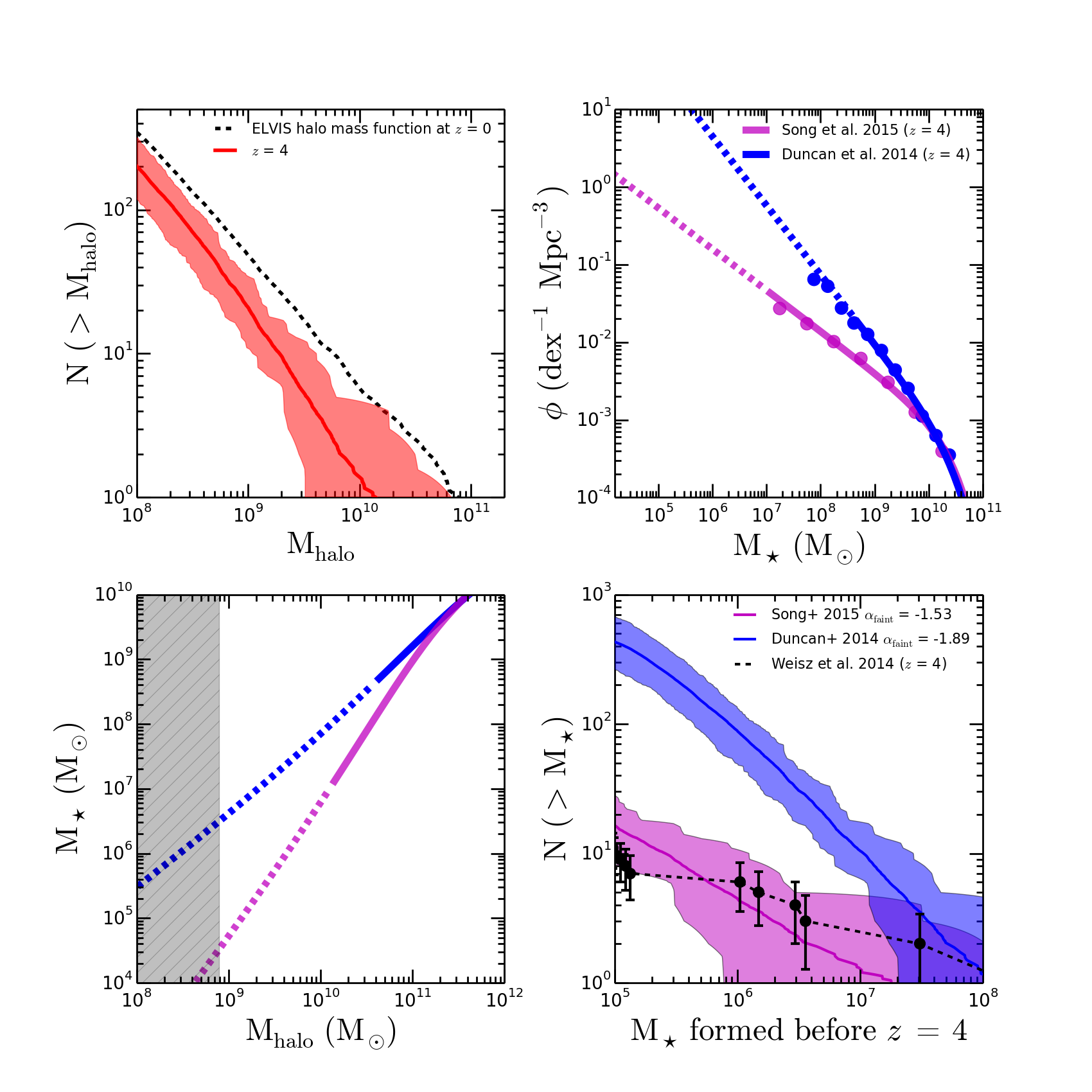}
\caption[Telescope-specific 3-color images]{\textit{Top left:} Cumulative halo mass functions for all bound objects within 300 kpc of simulated Milky Way analogs from {\tt ELVIS}.  The black dashed line shows the average mass function at $z=0$.  The red line shows the average mass function for main branch halos at $z = 4$ that survive as independent halos to the present day.  The shaded band corresponds to the full scatter over all 20 systems.  \textit{Top right}: GSMFs from D14 and S14, along with their published Schechter fits extrapolated down to the stellar masses of relevance for this paper.  \textit{Bottom left}: The stellar mass vs. halo mass relations that result from the same pair of GSMFs at $z=4$.   The shaded grey band in the lower left panel corresponds to the halo mass scale where galaxy formation may be suppressed by the UV background \citep{Okamoto08}. \textit{Bottom right}: The implied cumulative stellar mass functions before $z = 4$ that result from combining the upper left and bottom left panels.  The data points are measurements for Milky Way satellites from W13 and W14 with Poisson error bars. We see that an extrapolation of the D14 GSMF would drastically over-produce the satellite population of the Milky Way.  The only way the D14 normalization can be compatible is if the GSMF truncates sharply below $\Mstar \sim 10^7$ $\Msun$.  The S15 extrapolation, with its flatter faint-end slope, is remarkably consistent with satellite counts around the Milky Way.}
\label{f:4array}
\end{figure*}

\subsubsection{High redshift stellar mass functions}

We will focus our comparisons to measurements of high-redshift GSMF presented over the last two years.   For our $z=2-3$ constraints we will normalize at the high-mass end to the results of \citet[][I13 herafter]{Ilbert13}, who have used UltraVISTA to measure the GSMF to $M_\star  \gtrsim 10^{10}$ $\Msun$; and also to those of \citet[][T14 hereafter]{Tomczak14}, who have pushed a factor of $\sim 2$ deeper at the relevant redshifts using  ZFOURGE/CANDELS.

At $z=4-5$,  we will use the \textit{HST}-CANDELS + \textit{Spitzer}/IRAC results from \citet[][D14 hereafter]{Duncan14} and the more recent (deeper) study from \citet[][S15 hereafter]{Song15}, who reach $M_\star \simeq 10^7$ $\Msun$ (vs. $10^8$ $\Msun$ for D14) at these redshifts.  \citet{Grazian15} have presented a similar analysis and their fits are intermediate between those of D14 and S15; we focus on the latter two in our comparison in order to bracket the most recent results.

\begin{figure*}
\includegraphics[width=3.3in, height=3.3 in, trim = .2in 0 0 0]{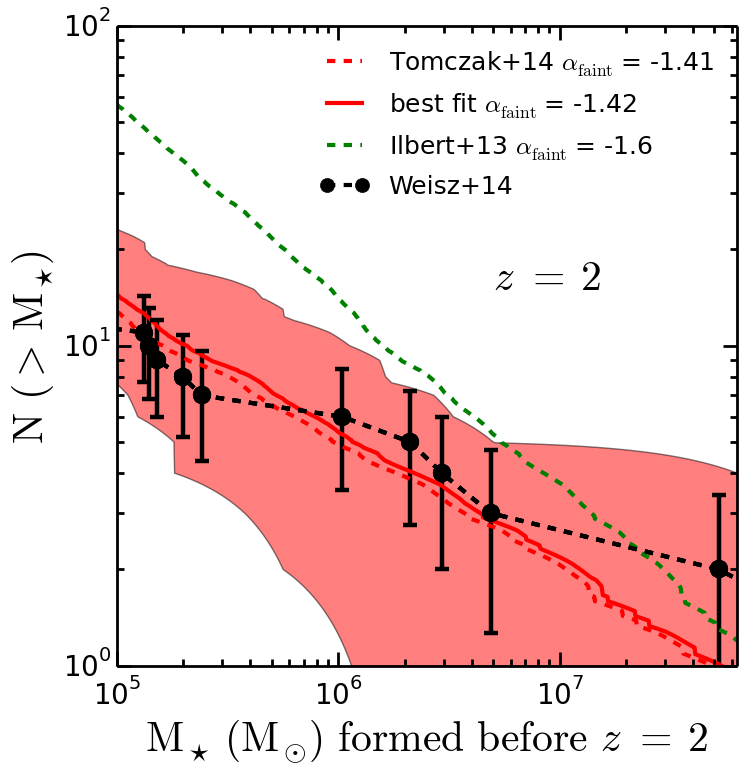}
\hspace{.2in}
\includegraphics[width=3.3in, height=3.3 in, trim = .2in 0 0 0]{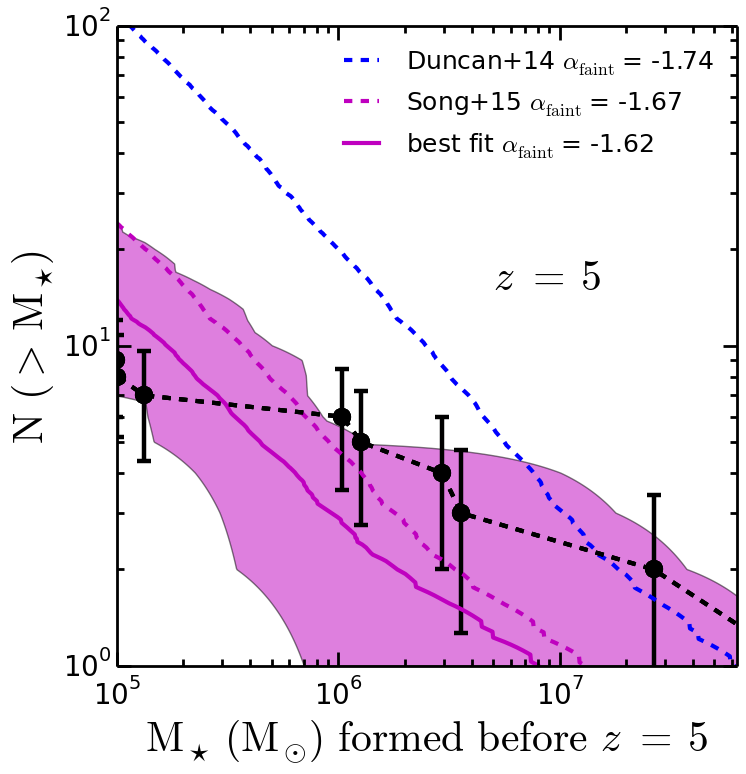}
\centering
\caption[Telescope-specific 3-color images]{Projections of the stellar mass function at low masses from $z=2$ to $z=5$ plotted along with present day measurements of the stellar mass function from various sources. The solid lines are best fit faint end slopes derived by fixing the other Schechter parameters to literature values and performing a chi-squared minimization. In general the best fit faint end slope does not differ much from those in the literature, and the faint end slope appears to be consistently near -1.5 for our redshift range.}
\label{f:4array}
\end{figure*}

\section{Motivation and Methodology} \label{s:discussion}

Figure 1 provides an overview of the strategy of this work.  The black dashed line in the upper-left panel shows the average mass function of bound halos within 300 kpc of the {\tt ELVIS} hosts at $z=0$ (hereafter called subhalos).   The red line shows the mass function of each bound subhalo's {\em main progenitor} at $z=4$.  The shaded region shows the full scatter in progenitor mass functions over all 20 {\tt ELVIS} hosts.  We see that each Milky Way host should have $\sim 200$ bound halos within 300 kpc that had progenitors more massive than $10^8$ $\Msun$ at $z=4$.   Given that the number of satellite galaxies of the Milky Way with $M_\star \gtrsim 10^5$ $\Msun$ is significantly lower than this ($12$), we know that any GSMF that requires $10^8$ $\Msun$ halos to host $M_\star \gtrsim 10^5$ $\Msun$ galaxies at $z=4$ would be inconsistent with known properties of the Milky Way.  The aim of this work is to use this basic tension to inform our understanding of the on the low-mass end of the GSMF at high redshifts, place limits on the asymptotic slope $\alpha_{\rm faint}$.

The upper right panel of Figure 1 compares the best-fit GSMFs of D14 (with $\alpha_{\rm faint} = -1.89$) and S15 (with $\alpha_{\rm faint} = -1.53$) at $z=4$.  Table 1 lists the preferred Schechter parameters from D14 and S15; the lines in Figure 1 show these fits and become dashed beyond the last published data point.   

Note that the comoving number density of $M_\star = 10^5$ $\Msun$ galaxies in the D14 extrapolation exceeds $10$ Mpc$^{-3}$.  We can  infer right away that such an extrapolation is problematic, as only halos with virial masses below $10^8$ $\Msun$ at these epochs have number densities this high \citep[e.g.,][]{Lukic07}.  That is, the only way to have a galaxy population this abundant at $z=4$ is to associate them with halos of mass $\le 10^8$ $\Msun$.  As we have just discussed, the $z=0$ descendants of such galaxies would drastically over-populate the Milky Way's local volume.

To add some precision to this constraint we rely on abundance matching, and specifically assume that there is a one-to-one relationship between the most massive halos and the most massive galaxies at $z=4$ (we discuss potential caveats with this assumption in \S 5).  The lower-left panel of Figure 1 shows the implied halo mass / stellar mass  relation for both GSMFs.  We have used the \citet{Sheth01} mass function for halo abundances here \footnote{and rely on the hmf {\tt Python} package \citep{Murray13}.}.  These relations allow us to assign a stellar mass to each progenitor subhalo mass at $z=4$ and then to construct a prediction for satellite stellar mass functions restricted to stars formed before $z=4$.   

The culmination of this exercise is shown in the bottom right panel of Figure 1.  The black points show the cumulative function of galaxies within 300 kpc of the Milky Way as a function of $M_\star^{z=4}$ -- their mass in ancient stars formed prior to $z=4$ (from W13, W14).
We note that the overall count is low ($10$ with $M_\star ^{z=4} > 10^5$ $\Msun$) and that the mass function is fairly flat $N \propto (M_\star^{z=4})^{1/3}$. These counts are in stark contrast to what would be expected if an extrapolation of the D14 GSMF were to hold at $z=4$ (blue), as this would give a steeply rising mass function ($N \propto M_\star^{z=4}$) and result in $> 400$ galaxies with $M_\star^{z=4} > 10^5$ $\Msun$.  The shading corresponds to the full halo-to-halo scatter in our simulations and the solid line shows the median.  If instead we assume the S15 fit holds (magenta), the result is much more in line with the Milky Way data.  Another possibility (not shown) is that the D14 normalization is correct but that the stellar mass function at $z=4$ sharply breaks at $M_\star < 10^7$ $\Msun$ so that the number density at $M_\star \sim 10^5$ $\Msun$ is something closer to the S15 extrapolation in the upper right panel. 

%%%%%%%%%%%%%%%%%%%%%%%%%%%%%%%%%%%%%%%%%%%%%%%%%%%%%%%%%%%%%%%
\begin{table*}
  \caption{
	Schechter parameters for the reference stellar mass functions used in this work, along with the calculated best fit faint end slopes using local group data.  Literature stellar mass functions from Ilbert+ 13 and Tomczak+ 14 were fit using a double Schechter function. Note that at $z=2$ the literature fits for Ilbert+ 13 and Tomczak+14 are actually two separate bins (1.5 $<z<$ 2.0 and 2.0 $<z<$ 2.5) which we simply averaged to get the parameters listed here.  The same procedure was used at $z=3$ for the Ilbert+ 13 data (using bins 2.5 $<z<$ 3.0 and 3.0 $<z<$ 3.5) and for Tomczak+ 14 we used the last data bin (2.5 $<z<$ 3.0). Rows 2 and 5 through 8 are the literature values of the Schechter fit, while Rows 3 and 4 represent the best fit slopes using local group data and keeping the rest of the Schechter fit paramters fixed to their literature values.
    \label{table:table1}
  }
  \begin{tabular*}{0.94\textwidth}{@{\extracolsep{\fill} } ccccccccc}
    \hline
    \\
    \multicolumn{5}{l}{Normalized to I13 (UltraVISTA DR 1)}\\
    \hline
      & $z$	&	$\alpha_{\rm faint}$	&	$\alpha_{\rm faint}$	&	$\alpha_{\rm faint}$	&	$\log(M^{*})$	&	$\Phi^{*}_{1}$	&	$\alpha_{1}$	&	$\log(\Phi^{*}_{\rm faint})$\\
      &		 &	(reference)	&	(cumulative fit)	&	(differential fit)	&	$(M_{\odot})$		&	$(10^{-3} {\rm Mpc}^{-3})$&		&	$(10^{-3} {\rm Mpc}^{-3})$\\
   
	\hline
	  & 2	&	\big[ -1.60 \big]	&	$-1.48_{-0.03}^{+0.05}$	&	$-1.43_{-0.03}^{+0.06}$	&	$10.74$	&	$0.750$	&	$-0.23$	&	$0.240$\\
	  &	3	&	\big[ -1.60 \big] &	$-1.58_{-0.04}^{+0.03}$	&	$-1.55_{-0.03}^{+0.06}$	&	$10.75$	&	$0.145$	&	$0.40$	&	$0.145$\\
	\hline
	\\
	\multicolumn{5}{l}{Normalized to T14 (ZFOURGE/CANDELS)}\\
	\hline
	  & 2	&	\big[$-1.41_{-0.22}^{+0.26}$ \big]	&	$-1.42_{-0.03}^{+0.06}$	&	$-1.37_{-0.04}^{+0.06}$	&	$10.715$	&	$0.375$	&	$0.535$	&	$0.479$\\
	  & 3	&	\big[$-1.57_{-0.20}^{+0.20}$ \big]	&	$-1.50_{-0.05}^{+0.02}$	&	$-1.49_{-0.03}^{+0.07}$	&	$10.74$	&	$0.029$	&	$1.62$	&	$0.204$\\	
	\hline
	\end{tabular*}
\end{table*}
%%%%%%%%%%%%%%%%%%%%%%%%%%%%%%%%%%%%%%%%%%%%%%%%%%%%%%%%%%%%%%%
%%%%%%%%%%%%%%%%%%%%%%%%%%%%%%%%%%%%%%%%%%%%%%%%%%%%%%%%%%%% 
\begin{table*}
	\caption{
	Similar to Table 1 except at $z=4$ and $5$.  In this case, both literature reference use a single Schechter fit in place of the double Schechter fit used in the Table 1 references.  Our best fit slopes were calculated in the same manner, by keeping $M*$ and $\Phi*$ fixed and leaving $\alpha$ as a free parameter. 
	\label{table:table1}
	  }
	 \begin{tabular*}{0.94\textwidth}{@{\extracolsep{\fill} } ccccccccc}
	 \hline
	 \\
	 \multicolumn{5}{l}{Normalized to S15 (CANDELS/HUDF)}\\
	 \hline
	 & $z$	&	$\alpha_{\rm faint}$	&	$\alpha_{\rm faint}$	&	$\alpha_{\rm faint}$	&	$\log(M^{*})$	&	$\log(\Phi^{*})$	\\
	 & 	&	(reference)			&	(cumulative fit)	&	(differential fit)	&	$(M_{\odot})$	&	$(10^{-3} {\rm Mpc}^{-3})$\\  
	 \hline
	 & 4	&	\big[$ -1.53_{-0.06}^{+0.07}$ \big]	&	$-1.52_{-0.04}^{+0.03}$	&	$-1.48_{-0.04}^{+0.06}$	&	$10.44$	&	$0.301$\\
	 &	5	&	\big[$ -1.67_{-0.07}^{+0.08}$ \big]	&	$-1.62_{-0.03}^{+0.04}$	&	$-1.55_{-0.06}^{+0.04}$	&	$10.47$	&	$0.134$\\
	 \hline
	 \\
	 \multicolumn{5}{l}{Normalized to D14 (CANDELS)}\\
	 \hline
	 & 4	&	\big[ $-1.89_{-0.13}^{+0.15}$ \big]	&	$-1.56_{-0.04}^{+0.03}$	&	$-1.52_{-0.04}^{+0.06}$	&	$10.54$	&	$0.189$\\
	 & 5	&	\big[ $-1.74_{-0.29}^{+0.41}$ \big]	&	$-1.59_{-0.03}^{+0.04}$	&	$-1.50_{-0.05}^{+0.04}$	&	$10.68$	&	$0.124$\\	
	 \hline
	 \end{tabular*}
\end{table*}
%%%%%%%%%%%%%%%%%%%%%%%%%%%%%%%%%%%%%%%%%%%%%%%%%%%%%%%%%%%%%%%

\section{Constraints}

Figure 2 presents the results of a similar exercise to that performed in the previous section but now at $z=2$ (left) and $z=5$ (right).  As in the lower right panel of Figure 1, the black points show the cumulative stellar mass functions for Milky Way satellites within 300 kpc, specifically the stellar mass formed prior to $z=2$ (left, $M_\star^{z=2}$) and $z=5$ (right, $M_\star^{z=5}$).  

The dashed lines in the left panel of Figure 2 shows the implied (median) cumulative mass function, $N(>M_\star^{z=2})$, for the fiducial GSMF at $z=2$ from I13 (green dash) and T15 (red dash).  Note that neither of these data sets were able to reach down to the low stellar masses shown in the plot; the results presented correspond to extrapolations of their double-Schechter fits (with parameters listed in Table 1).  Note that I13, motivated by their fits at lower redshift, have fixed $\alpha_{\rm faint} = -1.6$ in their fiducial GSMFs.  If this were to hold true, the left-hand side of Figure 2 shows that this would predict typically $\sim 55$ galaxies within 300 kpc of the Milky Way with more than $10^5$ $\Msun$ in stars older than about $10.4$ Gyr ($z=2$ formation).  This is about a factor of five more than observed, which is larger than the halo-to-halo scatter expected from the {\tt ELVIS} suite (shaded red band, see below).  The T14 line, on the other hand, agrees fairly well with the Milky Way data, owing primarily to its flatter slope, $\alpha_{\rm faint} = -1.41$.  

The red solid line and associated shaded band in Figure 2 shows the preferred fit for $\alpha_{\rm faint}$ ($-1.42$) that results from fixing the other parameters of the T14 GSMF and then fitting the implied bound descendants to the cumulative $M_\star^{z=2}$ data from Milky Way satellites.  The result is almost indistinguishable from the mass function measured by T14 at $z=2$.  This is remarkable given that we are performing a consistency check $\sim 10$ Gyr after the epoch observed and $\sim 3$ orders of magnitude lower in stellar mass than the limit of their survey.  If instead we do the fit to the differential mass function we find a preferred value $\alpha_{\rm faint} = -1.37$ for the T14 normalization.

The right panel of Figure 2 shifts the focus to $z=5$, where the dashed lines show the implied Milky Way mass functions in stars formed prior to this redshift assuming the D14 (blue dashed) and S15 (magenta dashed) GSMFs.  The D14 fit, with a steep slope $\alpha_{\rm faint} = -1.74$, over-predicts the known count by an order of magnitude.  The S15 GSMF at $z=5$ is much more consistent with what we see locally.  The solid magenta line and associated shading (full halo-to-halo scatter) shows the resultant best-fit $\alpha_{\rm faint}$ ($=-1.62$) when normalized to the other Schechter parameters in S15 (see Table 1).  The result is similar to the preferred value in S15 ($-1.67$), though slightly flatter.  Given the significant host-to-host scatter, the difference between our best-fit $\alpha_{\rm faint}$ and the S15 value is not significant, as we now discuss.  

Table 1 lists our best-fit $\alpha_{\rm faint}$ values when normalized to I13 and T14 at $z=2$ and $z=3$ for both cumulative mass function fits (column 2) and differential mass function fits (column 3). Table 2 lists our best-fit $\alpha_{\rm faint}$ values when normalized to D14 and S15 at $z=4$ and $z=5$.  The differential fits weight the existence of the (likely rare) Magellanic Clouds less than do the cumulative fits and therefore typically prefer slightly flatter slopes (at the $0.03-0.05$ level).  For all of the fits we perform we use the average of the cumulative stellar mass functions derived from {\tt ELVIS} to compare to the W14 sample in order to compute the best fit faint end slope.  We also compute errors on the faint end slope by fitting to the upper and lower 68 \% range of the {\tt ELVIS} hosts (e.g., consider the upper and lower ranges of the red band plotted in the upper left panel of Figure 1).  This provides an accounting of the expected cosmic variance in subhalo/satellite counts from halo-to-halo.  

Figure 3 summarizes our constraints on $\alpha_{\rm faint}$ as a function of redshift and compares our values to recent measurements reported in the literature.  We specifically plot the average of the four best-fit $\alpha_{\rm faint}$ values listed in Tables 1 and 2, with errors that treat variance among those four preferred values as systematic combined with the halo-to-halo variance errors listed for each of the four fits.   The specific values we prefer are: $\alpha_{\rm faint} = $ ($-1.42^{+0.07}_{-0.05}$, $-1.53^{+0.06}_{-0.05}$, $-1.52^{+0.05}_{-0.05}$, $-1.57^{+0.06}_{-0.06}$) at $z = $ ($2$, $3$, $4$, $5$).

\begin{figure*}
\centering
\includegraphics[width=3.0in, height=3.0in, trim = .8in 0 0 0 0]{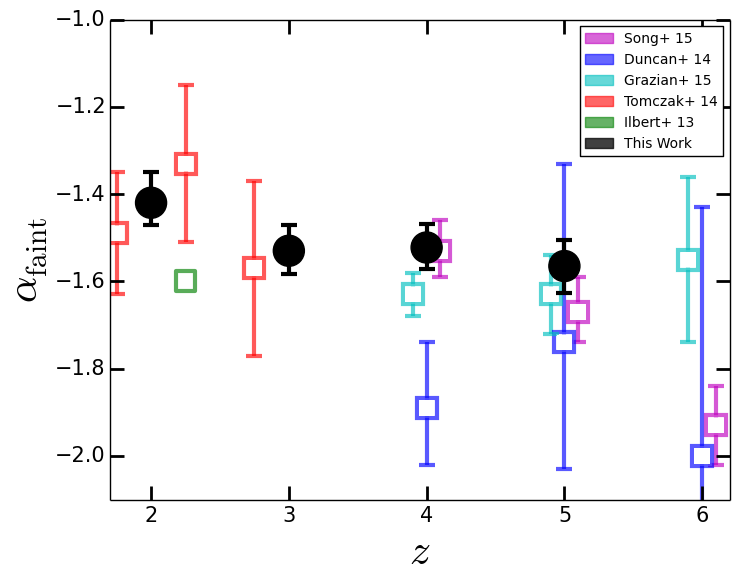}
\caption{Our preferred fits for  $\alpha_{\rm faint}$ as a function of redshift (black points) along with recent measurements of $\alpha_{\rm faint}$ from various direct-imaging surveys.  The redshifts of the \citet{Song15} and  \citet{Grazian15} results have been artificially offset for clarity. We include only one point from Ilbert et al. (2013) at $z \simeq 2$  because the faint end slope they adopt is fixed to -1.6 based of their results at $z = 1.5$.  Our black points were derived by simply averaging the best fit slopes from Tables 1 and 2 at each redshift.  Our error bars include the variance among those four slopes as well as an accounting of halo-to-halo scatter in subhalo abundances.}
\label{f:4array}
\end{figure*}

\section{Discussion}

In this paper we have used near-field observations (deep CMD-based SFHs from W13 and W14) together with $\Lambda$CDM simulations to inform our knowledge of the galaxy stellar mass function (GSMF) at early times ($z=2-5$).  The power of this approach is that it utilizes a huge lever arm in stellar mass,  exploring a mass regime two orders of magnitude below what is currently possible via direct counts. Our results suggest that the low-mass slope of the GSMF must remain relatively flat with only slight steepening over the redshift range $z=2-5$, from $\alpha_{\rm faint}^{z=2} \simeq   -1.42^{+0.07}_{-0.05}$ to $\alpha_{\rm faint}^{z=5} \simeq -1.57^{+0.06}_{-0.06}$ at $z=5$.    These constraints  are strongly at odds with some published GSMFs, including published estimates that would over-produce local galaxy counts by an order of magnitude or more (see, e.g., the lower right panel of Figure 1).

While the discriminating capacity of this approach is potentially very powerful, there are a few caveats that we must consider.  The first is the concern that our simulations may not provide a fair representation of the Milky Way and its local environment.  The {\tt ELVIS} hosts we use have virial masses that vary over the range $M_{\rm halo} = 1-2.4 \times 10^{12}$ $\Msun$.  This spans the range of most recent measurements but there are some estimates that fall as low as $7 \times 10^{11}$ $\Msun$ \citep[see, e.g.][]{MBK13,Wang15}.  If the Milky Way halo mass is really this low, then we would expect roughly a factor of $\sim 2$ fewer progenitors at fixed halo mass within $300$ kpc of the Milky Way.  This would shift our median predictions for bound satellite counts down by roughly the same factor for any given GSMF, which is within the shaded regions shown, e.g., in Figure 2.  Our preferred faint-end slopes would similarly be systematically steeper, roughly $\alpha_{\rm faint} \rightarrow \alpha_{\rm faint} - 0.05$.  Such a shift would not change our conclusion that the preferred low-mass behavior of the I13 and D14 GSMFs are too steep to be compatible with local counts and would not change our conclusions that $\alpha_{\rm faint}$ is evolving weakly to $z=5$.    

A similar concern is associated with the fact that our simulations are pure N-body and lack the additional gravitational potential associated with a central disk.  The extra tidal forces associated with the disk will likely liberate mass from some subhalos such that our N-body runs over-predict bound subhalo counts \citep[][]{Donghia10,Brooks14}.  We suspect that this effect will marginally affect our work because we count bound structures based on their {\rm progenitor} masses prior to accretion, not on their masses at $z=0$ (which will certainly be lower than predicted from N-body for subhalos that orbit close enough to the disk).  Progenitor masses will be
largely unaffected by the presence of a disk.  A more important concern is that more satellites will be completely destroyed when the disk potential is included, thus liberating their stars to the stellar halo \citep[e.g.,][]{BJ05}.  Given that the total stellar mass of the Milky Way's stellar halo is quite modest ($\sim 10^9$ $\Msun$) compared to the integrated mass of the stellar mass functions shown in, e.g., Figure 2, there is not much room to hide additional mass in tidally destroyed galaxies that are not already accounted for in current N-body based models \citep{Lowing15}.  More work needs to be done in order to solidify this expectation \citep[see, e.g.][for a promising approach]{Brooks13}.

Another issue that could affect our results is that we have chosen to use one-to-one abundance matching at $z=2-5$ in order to assign ancient stellar masses to the progenitors of our bound subhalos at $z=0$.  Imposing significant scatter in the $M_\star - M_{\rm halo}$ relation could drastically affect our constraints.  By definition, any imposed scatter would result in a GSMF that is identical (on average) to those used in the paper; once scatter is imposed, the median $M_\star - M_{\rm halo}$ needs to become steeper in order to result in the same overall number density (e.g., Garrison-Kimmel et al, in preparation).  This means that the {\em median} constraints on $\alpha_{\rm faint}$ would remain the same, but the uncertainty caused by host-to-host scatter would increase accordingly. In order to better understand the effect scatter would have on our results we performed a simple experiment, calculating cumulative stellar mass functions in the same manner, but allowing the stellar mass given by our abundance matching relations to scatter by 0.2 dex according to a log-normal distribution. The resulting best fit faint end slope shifts as anticipated, however the shift is small (about $\pm$ 0.01) so we believe our results are robust unless the scatter in abundance matching is much larger than 0.2 dex.

 While a number potential caveats need to be understood before we can gain further accuracy and precision on our constraints, this paper (along with B14 and B15) has illustrated the potential power of near-field data sets in constraining science that is traditionally relegated to deep field approaches.  None of the concerns we have discussed can account for the discrepancy between Milky Way satellite counts and the D14 GSMF at $z=4$ (see Figure 1); a mass function as steep as $\alpha_{\rm faint} \simeq -1.9$ at $z=4$ is effectively impossible to reconcile with our current understanding of structure formation.  
 
There are other lines of evidence from the near-field that can inform our understanding of the relationship between halo mass and stellar mass (and thus galaxy abundances) at high redshift.  For example, it has been known for some time that reionization will likely shut down galaxy formation below a critical halo mass scale of about $\sim 10^9 \Msun$ \citep{Efstatiou92,Shapiro94} and that this might leave imprints on the count and star-formation histories of Local Group galaxies that reside within those small halos \citep{Bullock00,Ricotti05, Gnedin06, Bovill09}.  The shaded grey band in the lower left panel of Figure 1 corresponds to the halo mass scale where galaxy formation is expected to be suppressed by the external UV ionizing flux at $z=4$ according to \citet{Okamoto08}.  The D14 abundance-matching relation would suggest that we should begin seeing the effects of reionization suppression in galaxies as massive as $M_{\star} = 10^6$ $\Msun$ at $z=4$; however, no systematic truncation in star formation at this epoch is seen in the CMD-derived SFHs of galaxies this massive (W13, W14).  The steeper S15 relation, on the other hand, would suggest that UV suppression sets in at the stellar mass regime of ultra-faint dwarfs ($M_\star \sim 10^4$  $\Msun$).  This expectation, which only holds for flatter stellar mass functions as preferred by S15 (and our own work), agrees very well with recent observations that have seen uniformly old stellar populations in ultra-faint galaxies \citep{Brown14} and recent simulations \citep{Onorbe15,Wheeler15} as well.  More data will be needed to confirm whether this dividing line between uniformly ancient stellar populations and continued star formation is sharp at the stellar mass scale of ultra-faint dwarfs, but it is clear that the SFHs of local dwarf galaxies have a lot to tell us about the physics of galaxy formation at early times.

Finally, it is worth stressing the major assumption of this work, and the general concept of using the Local Group as a time machine, which is that the galaxies of the Local Group filled a much larger volume at high redshift, and thus are a representative sample of the dwarf population as a whole at high redshift. It is possible that this assumption is not correct and that dwarf galaxies that are destined to fall into the Local Group are somehow biased with respect to the bulk of the population at high redshift.  Future direct surveys of the high-redshift universe that allow deeper constraints on $\alpha_{\rm faint}$ will provide a test of this idea. If future counts measure different values of $\alpha_{\rm faint}$ than we anticipate from near-field studies, then this would help us understand how (and why) Local Group dwarfs would be biased relative to the general population of dwarf galaxies -- a result that would be important (if true). 

The approach adopted here will be further strengthened by more complete observational surveys of the Local Group.  In the near term, complete SFHs from M31 dwarfs will be invaluable.   Andromeda is believed to have a similar halo mass to the Milky Way, so including an additional constraint from this system will help us account for halo-to-halo scatter in subhalo populations, which dominates the error we report on $\alpha_{\rm faint}$.   More generally, with surveys like LSST poised to come on line, the next decade will almost certainly see the discovery of many more dwarf galaxies throughout the Local Volume, pushing our completeness limits to lower stellar masses and more distant radii \citep{Tollerud08}.  These discoveries, and associated follow-up, will further lengthen the lever arm on stellar mass functions and increase the volumes within which these constraints can be applied, thus decreasing cosmic variance uncertainties.  The same instruments that promise to reveal the high redshift universe directly (e.g., {\textit{JWST}}, TMT, GMT, E-ELT) will also facilitate indirect constraints on these epochs via archeological studies in the very local universe.  

\section*{Acknowledgements}
AGS was supported by an AGEP-GRS supplement to NSF grant AST-1009973.  MBK acknowledges support provided by NASA through Hubble Space Telescope theory grants (programs AR-12836 and AR-13888) from the Space Telescope Institute (STScI), which is operated by the Association of Universities for Research in Astronomy (AURA), Inc., under NASA contract NAS5-26555. DRW is supported by NASA through Hubble Fellowship grant HST-HF-51331.01 awarded by the Space Telescope Science Institute.  JSB was supported by HST AR-12836 and NSF AST-1009973.

\clearpage

\end{document}